\documentstyle[amssymb,aps,twocolumn]{revtex}

\begin{document}

\twocolumn[\hsize\textwidth\columnwidth\hsize\csname@twocolumnfalse\endcsname
]

{\bf Bobroff et al. Reply to the comment of Tallon{\it \ et al.},
Cond-mat/0008295:} The existence of local moments induced by spinless
impurities in cuprates has often been denied by Tallon and coauthors, e.g. 
\cite{Williams}.\ In \cite{comment}, they finally accept our interpretation
of the $^{7}$Li NMR shifts in terms of induced local moments nearby Li in
the underdoped and optimally doped regime. In contrast they still maintain
that our conclusion is incorrect in the overdoped regime. This issue is
certainly important in their scenario for which the abrupt change of
physical properties (absence of the pseudo gap) which occurs near $p=0.19$
would correspond to the disappearance of magnetic correlations. \ Their
argument is based on the {\em qualitative} analogy between the $T$
dependences of their $^{89}$Y\ and our $^{7}$Li shifts.\ This leads them to
attribute this $T$ variation in the overdoped regime to the intrinsic bulk
susceptibility $\chi $ of the CuO$_{2}$\ planes.\ In Fig. 1 we show that
apart the variation of the Kondo temperature, the $T$ dependences of the $%
^{7}$Li NMR\ shift in the YBa$_{2}$Cu$_{3}$O$_{7}$ and Y$_{1-x}$Ca$_{x}$Ba$%
_{2}$Cu$_{3}$O$_{7}$ substituted samples of \cite{BobroffPRLLi} have similar
magnitudes.\ These data cannot therefore be associated with two totally
different physical effects, a local moment in one case and the pure material 
$\chi $ in the other. The failure of this qualitative argument is even more
obvious from the following simple {\em quantitative} comparison.\ 

$^{7}$Li and $^{89}$Y nuclei are coupled to the CuO$_{2}$ planes
respectively through their 4 and 8 n.n. coppers.\ Hence, if $^{7}$Li and $%
^{89}$Y nuclei probe the uniform $\chi $ of the pure material their shifts
should be related by 
\begin{equation}
^{7}K-^{7}K_{0}=\frac{1}{2}\frac{^{7}A_{hf}}{^{89}A_{hf}}(^{89}K-^{89}K_{0})
\label{equ}
\end{equation}
where $^{7}A_{hf}=2.45kOe/\mu _{B}$ ,$^{7}K_{0}=90ppm$ \cite{BobroffPRLLi}
and $^{89}A_{hf}=-2kOe/\mu _{B}$, $^{89}K_{0}=150ppm$ \cite{alloul89} are
the hyperfine couplings between the $^{7}$Li and $^{89}$Y nuclei and one Cu
of the CuO$_{2}$ planes, and the reference shifts. These parameters do not
change significantly over the entire phase diagram and can be safely used in
the overdoped regime. From the data of Fig.1 of the comment we compute,
using Eq.(\ref{equ}), the expected $^{7}$Li shifts in Tallon's scenario.\
They exhibit much weaker $T$ variations than the actual data even for the
largest overdoping given in \cite{comment}. The $T$ variation of the $^{7}$%
Li shift is thus much too large to be attributed to the bulk $\chi $, even
for $p\geqslant 0.19$, for which the $^{89}$Y\ satellites cannot be
resolved. As $\chi $ has a large $T$ independent component in the pure
samples, fitting both the magnitude of the $T$ variation and the actual high 
$T$ limiting value for $^{7}K$ is impossible for any value of $%
^{89}A_{hf}/^{7}A_{hf}$.\ Rather, the continuous variation of $^{7}K$ and $%
^{7}T_{1}$ \cite{AMF} with hole doping \ supports our interpretation in
terms of induced local moments which are Kondo-screened by the carriers
introduced by doping. This implies, contrary to the speculation of \cite
{comment}, that the magnetic correlations responsible for the existence of
such moments still play a role in the overdoped regime.

The other point considered ''less secure'' in \cite{comment} concerns the
connection between the electronic properties of the impurities and the
scattering.\ We agree that one cannot infer the physical origin of the
scattering solely from comparison of the $T_{c}$ variations induced by Ni
and Zn (or Li) . However the data taken in our group clearly indicate that
these two types of impurities display totally different {\em magnetic}
properties both of which reveal the occurence of magnetic correlations in
the host\ (the local spin S=1/2 on the Ni site is smaller than the expected
S=1 value\ \cite{mendels}). This remarkable difference has been recently
confirmed by STM measurements of the local electronic structure \cite{Pan}.
The correlation between the microscopic electronic structure and the
scattering is evidently not simple, and thus in \cite{BobroffPRLLi} we made
no attempt to speculate on this point.\ However, the anomalously large
scattering by the Zn defect has very often been attributed to pure local
potential scattering \cite{Bernhardt}.\ This is somewhat awkward as Zn$^{2+}$
and Cu$^{2+}$ have the same charge.\ We suggested that it is rather linked
with the existence of electronic correlations which result in a Kondo
resonance.\ This is strenghtened by the fact that the charge difference
between Zn and Li is not reflected both in their local magnetic properties
and the T$_{c}$ variation.

legend of fig.1 :
\begin{figure}[tbp]
\caption[1]{$^{7}$Li\ NMR\ shift data of Ref.\protect\cite{BobroffPRLLi}
compared with that expected from the pure samples susceptibility using Eq.(%
\ref{equ}) and the data from the comment. In Tallon et al.'s hypothesis,
solid (empty) circles should fit dash-dot (solid) curves respectively, which
correspond to similar hole dopings.}
\label{fig.4}
\end{figure}

J.\ Bobroff, W.A.\ MacFarlane, H.\ Alloul, P.\ Mendels

Laboratoire de Physique des Solides, UMR\ 8502, CNRS 91405 Orsay, France%
\newline

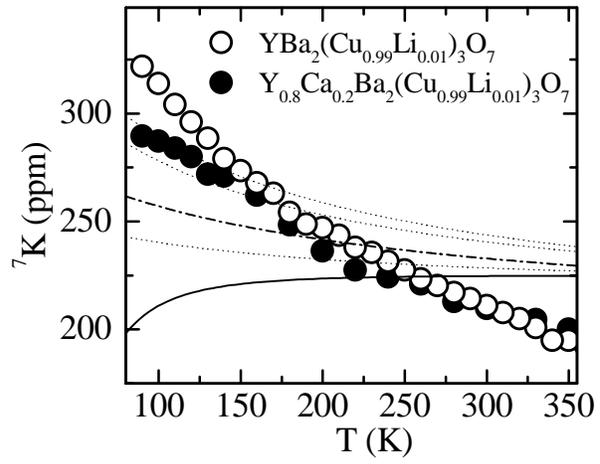
\end{document}